\documentclass[prl,twocolumn,showpacs,tightenlines,
preprintnumbers,amsmath,amssymb]{revtex4}

\usepackage{graphicx}
\newcommand{\beq}{\begin{equation}}
\newcommand{\eeq}{\end{equation}}
\newcommand{\bea}{\begin{eqnarray}}
\newcommand{\eea}{\end{eqnarray}}
\newcommand{\ba}{\begin{array}}
\newcommand{\ea}{\end{array}}
\newcommand{\bc}{\begin{center}}
\newcommand{\ec}{\end{center}}

\newcommand{\bml}{\begin{mathletters}}
\newcommand{\eml}{\end{mathletters}}
\newcommand{\commentout}[1]{{}}

\newcommand{\half}{\hbox{$1\over2$}}
\newcommand{\quarter}{\hbox{$1\over4$}}

\newcommand{\eq}[1]{(\ref{#1})}
\newcommand{\comment}[1]{{}}

\newcommand{\etal} {et al.}
\begin{document}
\title{Pulsating instability of a Bose-Einstein condensate in an optical lattice}
\author{Uttam Shrestha}
\author{Marijan Kostrun}
\author{Juha Javanainen}
\affiliation{Department of Physics, University of Connecticut,
Storrs, CT 06269-3046}
\date{\today}

\begin{abstract}
We find numerically that in the limit of weak atom-atom interactions a Bose-Einstein condensate in an optical lattice may develop a pulsating dynamical instability in which the atoms nearly periodically form a peak in the occupation numbers of the lattice sites, and then return to the unstable initial state. Multiple peaks behaving similarly may also occur. Simple arguments show that the pulsating instability is a remnant of integrability, and give a handle on the relevant physical scales.
\end{abstract}
\maketitle


Wave propagation combined with nonlinearity frequently gives rise to what is known as modulational or dynamical instability. It has been observed experimentally~\cite{STR02} that a one-dimensional Bose-Einstein condensate (BEC) with attractive interactions between the atoms is unstable, and may break up into a chain of solitons~\cite{RUP95}. Instabilities result from the interplay between dispersion and nonlinearity, and confinement of a condensate into an optical lattice modifies the dispersion so that the range of potential nonlinear phenomena is correspondingly broader~\cite{MOR06}. The flow of a BEC in an optical lattice is  dynamically unstable~\cite{WU012,SME02} at certain quasimomenta even for repulsive interactions. The empirical view is that, upon the instability, the BEC in a lattice evolves irregularly~\cite{FAL04}. However, one expects~\cite{KON02} and finds experimentally~\cite{EIE04} that the instability may again lead to soliton formation. Although our focus is on a BEC in an optical lattice, similar phenomenology has been studied in a wide range of nonlinear systems in many field of science. For instance, there is a precise analog of the experiment~\cite{EIE04} in nonlinear optics that employs a waveguide array~\cite{STE06}.

In this Letter we study a BEC in an optical lattice numerically for weaker atom-atom interactions than is customary~\cite{ROA07}. We find that the system may exhibit a pulsating dynamical instability in which the atoms nearly periodically collect to a peak in lattice occupation numbers, and subsequently disperse back to (very close to) the initial unstable state. Apparently related pulsations starting from an already compressed atom distribution in a lattice have been reported from numerical studies of the nonpolynomial Schr\"odinger equation~\cite{BAR07}, but the present framework affords many additional insights. By comparing with a two-site lattice, we see that, while the multisite lattice retains the instability as a vestige of the nonlinearity, it behaves approximately as if it were integrable when the instability unfolds. We are also able to develop simple quantitative estimates that delineate the conditions for the pulsating instability, and  correctly give the temporal and spatial scales of the pulsating atom distribution. 

The discrete nonlinear Schr\"odinger equation (DNLSE) serves as a tight-binding model for the amplitudes of the BECs $\alpha_k$ at the sites $k=0,\ldots,N-1$ of a one-dimensional optical lattice,
\beq
i\dot\alpha_k = -\half\delta(\alpha_{k+1}+\alpha_{k-1}) +\chi |\alpha_k|^2\alpha_k\,.
\label{EQM}
\eeq
We use periodic boundary conditions. The quantities $|\alpha_k|^2$ are proportional to the numbers of atoms at the sites $k$, and we refer to them as populations.  The constants $\delta$ and $\chi$ characterize tunneling of the atoms from site to site and on-site atom-atom interactions. Momentarily we discuss the case with $\delta>0$ and $\chi\ge0$. DNLSE is the equation of motion under the Hamiltonian
\beq
H =\sum_k\left[ -\half\delta\alpha^*_k(\alpha_{k+1}+\alpha_{k-1})
+\half\chi|\alpha_k|^4
\right]
\label{HAM}
\eeq
and the Poisson brackets $\{\alpha_k,\alpha^*_{k'}\} = -i\delta_{k,k'}$, so that the value of the Hamiltonian, the energy, is a constant of the motion. The normalization $\sum_k |\alpha_k|^2$ of the BEC amplitudes, which is proportional to the total number of atoms, is also a constant of the motion.  We normalize to the number of lattice sites $N$: $\sum_k |\alpha_k|^2 = N$. For notational simplicity $N$ is taken to be even.

The DNLSE admits stationary solutions of the form $\bar\alpha_k(p,t) = e^{i[pk-\mu(p) t]}$, with $\mu(p)=-\delta\cos p + \chi$. By virtue of the periodic boundary conditions the quasimomenta must be of the form $p=2\pi P/N$ with an integer $P$. In the reduced zone scheme all integers characterizing plane wave modes are folded into the interval  $(-N/2,N/2]$.
The standard ansatz for linear stability analysis for the plane-wave mode $p$ is
$\alpha_k(t;p,q)=e^{i[pk-\mu(p) t]}[u e^{i(q k-\omega t)}+  v^*e^{-i(q k-\omega^* t)}]$, 
where $u$ and $v$ are (small) constants. By the periodic boundary conditions the quasimomentum of an excitation $q$ (displacement from the quasimomentum of the stationary solution) must again be of the form $q=2\pi Q/N$ with an integer $Q\ne0$. This ansatz works, giving both the small-oscillation frequencies~\cite{SME02}
\bea
&&\omega(p,q) = \delta\sin p \sin q\nonumber\\
& \pm&\sqrt{
\delta\cos p(1-\cos q)[(2\chi + \delta\cos p(1-\cos q)]
}
\label{EXFRQS}
\eea
and the amplitudes $u$ and $v$ as a solution to an eigenvalue problem.

The expression inside the square root in Eq.~\eq{EXFRQS} is nonnegative for $p\in[-\half\pi,\half\pi]$ but may become negative otherwise, depending on the values of $\delta$, $\chi$, $p$ an $q$. Correspondingly, the square root may become imaginary, whereupon there are small deviation from the steady state that grow exponentially in time. The flow with the quasimomentum $p$ is then dynamically unstable. If there is an imaginary part to the eigenfrequency $\omega$, it is the same for excitation quasimomenta $q$ and $-q$. Moreover, the unstable direction, the eigenvector $[u,v]^T$ corresponding to the frequency with a negative imaginary part, is the same for $q$ and $-q$. In this sense the modes $q$ and $-q$ are equivalent as it comes to the instability.

\begin{figure}
\begin{center}
\includegraphics[width=8.5cm]{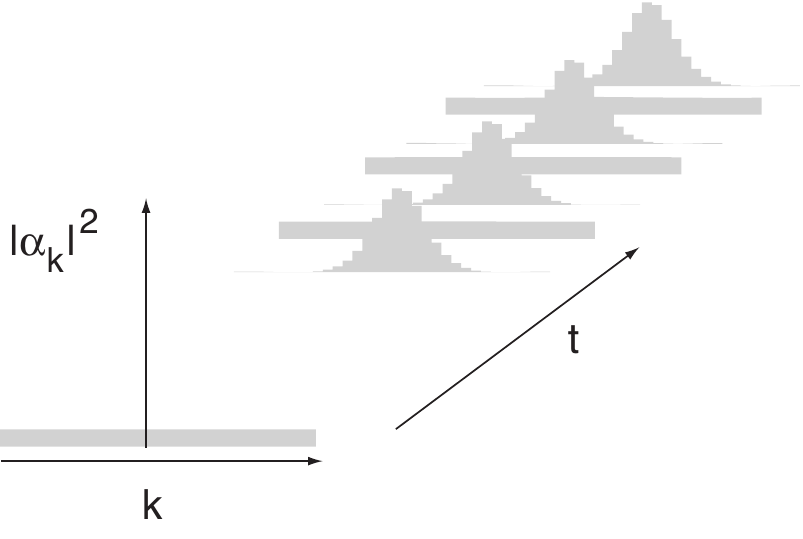}
\vspace{-20pt}
\end{center}
\caption{Snapshots of the population $|\alpha_k|^2$ at different times. The parameters are $N=32$, $\chi/\delta = 0.015$, and $p=\pi$. The instability is seeded with Gaussian noise with the amplitude $\xi=10^{-4}$.}
\vspace{-15pt}
\label{RECUR}    
\end{figure}

In our renewed look into the fate of an unstable mode we carry out numerical experiments on the DNLSE. Figure~\ref{RECUR} shows an example result. Here we have $N = 32$ lattice sites and an initial flow with $p=\pi$, or $P=N/2=16$. In this state $\bar\alpha_k(p,t)$ the sign of the condensate amplitude simply flips from one site to the next. The interaction is weak, $\chi/\delta=0.015$. To speed up the instability, 
we add complex-valued Gaussian noise to the amplitude of each site. Here the root-mean-square noise amplitude is $\xi=10^{-4}$. The Figure shows a histogram of the populations $|\alpha_k|^2$ at certain snapshot times. The instability leads to a single-peaked distribution of the occupation numbers. Moreover, upon further time evolution the system returns very close to the initial state, again pulsates to a peak, and so on. We have periodic peaking and recurrences of the unstable initial state.

\begin{figure}
\begin{center}
\includegraphics[width=8.5cm]{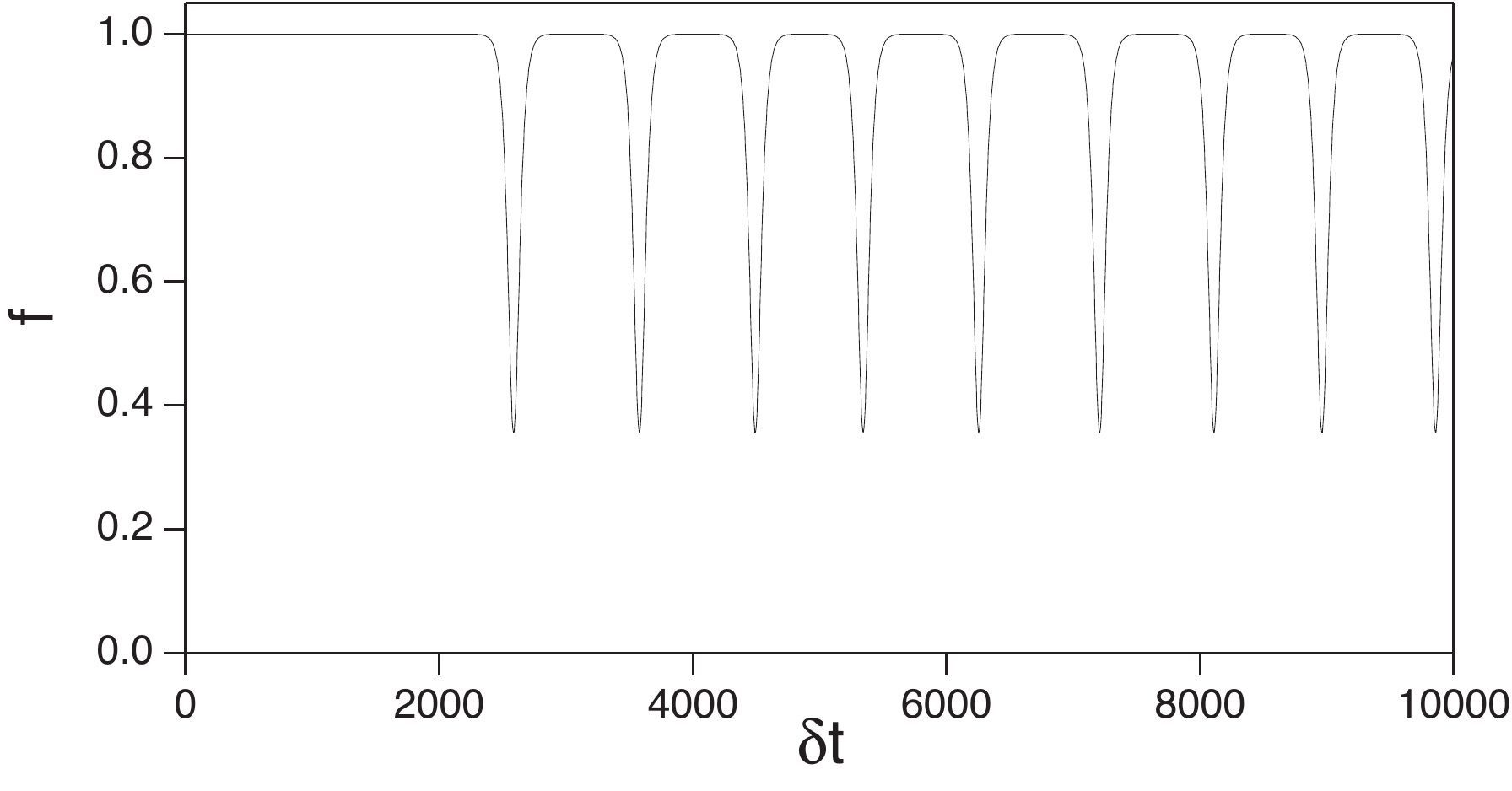}
\vspace{-20pt}
\end{center}
\caption{Fraction of the initial state in the state of the lattice, $f$, plotted as a function of time. This figure is from the same data as Fig.~\ref{RECUR}.}
\vspace{-15pt}
\label{PEAKCOMB}    
\end{figure}

Another view on the same data is given in Fig.~\ref{PEAKCOMB}. We plot the fraction of the initial state $\alpha_k(0)$ in the state at time $t$, $\alpha_k(t)$, the quantity $f(t) = \left| \sum_k \alpha^*_k(0)\alpha_k(t)\right|^2/N^2$, as a function of time. The times of the first four deep minima are the times picked for the snapshots of the four peaks in Fig.~\ref{RECUR}, but Fig.~\ref{PEAKCOMB} reveals five more peaking events. As appropriate for phenomena initiated by a dynamical instability, if everything else is held unchanged, the time of the first peak depends logarithmically on the amplitude of the noise $\xi$. A closer inspection reveals that the interval between the bottoms of the dips in Fig.~\ref{PEAKCOMB} varies slightly, so that the subsequent peaking events are not strictly periodic.

From data sets of this kind two observations emerge. First, if the peaking starts close to periodic, it may continue so seemingly indefinitely. We have examples with 50 recurrences. Second, near-periodic recurrences occur also for other initial flow states than the edge of the first Brillouin zone with $p=\pi$. In such a case the peaks recur at different positions in the lattice, and move at approximately the usual group velocity $v_g= \delta\sin p$.

The common predicate for quasiperiodic recurrences of a single peak is that, except for the $+Q$ and $-Q$ equivalence, one and only one of the possible modes $Q$ is unstable. In the case of more than a few lattice sites, this condition can only be satisfied for weak interactions, $\chi/\delta\ll1$, and the one possible unstable mode is $Q=1$. A straightforward analysis of the eigenvalues~\eq{EXFRQS} shows that the range of the interaction strength where the $Q=1$ mode is unstable and the $Q=2$ mode is not, in the limit $N\gg1$, is given by
\beq
|\cos p|({\pi}/{N})^2 <{\chi}/{\delta}<4|\cos p|({\pi}/{N})^2\,.
\label{PEAKCONDITION}
\eeq

A plot similar to Fig.~\ref{RECUR} shows that the Fourier transform of the site amplitudes $\alpha_k(t)$ with respect to the site index $k$ mainly displays the Fourier components $P$ and $P\pm1$. This is as should be, since the excitation modes that go unstable have the indices  $Q=\pm1$ with respect to the initial plane wave and therefore make  the Fourier components $P\pm1$. Depending on the parameters, the $P\pm1$ modes may almost completely extinguish the original mode $P$. The nonlinear dynamics also generates Fourier component $P\pm2$, etc., but with smaller amplitudes. If all atoms were in the plane wave modes $P\pm1$, we would have a modulation of the populations of the form $\cos^2(2\pi k/N+\varphi)$. This neatly explains why the width of the base of a recurring peak in the number of sites $k$ is about half of the size of the lattice $N$. In fact, let us use the lower limit of the interaction strength in Eq.~\eq{PEAKCONDITION} as a condition for the occurrence of a peak and the observation that the width of the peak is about $N/2$, then we have an estimate for the width of the peak as a function of the interaction strength $\chi$,
\beq
\ell\sim\pi\sqrt{\delta|\cos p|/\chi}\,/2\,.
\label{PEAKWIDTH}
\eeq
 We may then cast the condition for the occurrence of a single peak into the form that for a given interaction strength and the ensuing width of the peak $\ell$, one and only one peak comes up if and only if the length of the lattice is such that the lattice loosely accommodates one peak, but two peaks would have to overlap. 

Clean recurrences do not occur for small site numbers $>2$, such as $N=4$. Then a fairly strong nonlinearity $\chi/\delta\sim1$ is required for the instability, and many Fourier modes are coupled in. Besides, the Fourier components wrap around; $P+2$  and $P-2$ are the same plane wave mode for $N=4$. Apparently the nonlinearity strongly mixes all modes, which leads to irregular behavior.

In search of a qualitative explanation of the quasiperiodic behavior we study the system with just two sites. The transformation
$\alpha_{0,1} = e^{ 
i(\phi\pm\half\Delta\phi  )}\sqrt{\half n \pm \Delta n}$,
with $(n,\phi)$ and $(\Delta n,\Delta \phi)$ being usual canonical-conjugate pairs (Poisson brackets $\{n,\phi\}=1$, and so on) puts the two-site Hamiltonian into the form
\beq
H_2 = \chi(\quarter\,n^2 + \Delta n^2)-\delta\sqrt{n^2-4\Delta n^2}\cos\Delta\phi\,.
\label{TWOSITEH}
\eeq
Since there is no angle $\phi$ in the Hamiltonian, the canonical conjugate $n$, total normalization, is a constant of the motion, and $n=N=2$ according to our convention. $\Delta n$ and $\Delta\phi$ are, respectively, half of the population difference and the BEC phase difference between the sites 0 and 1. The potentially unstable steady state is  $P=1$, which in the present variables translates to $\Delta n=0$, $\Delta\phi=\pi$. The onset of the instability is at $\chi/\delta=1$.

\begin{figure}
\begin{center}
\includegraphics[width=8.5cm]{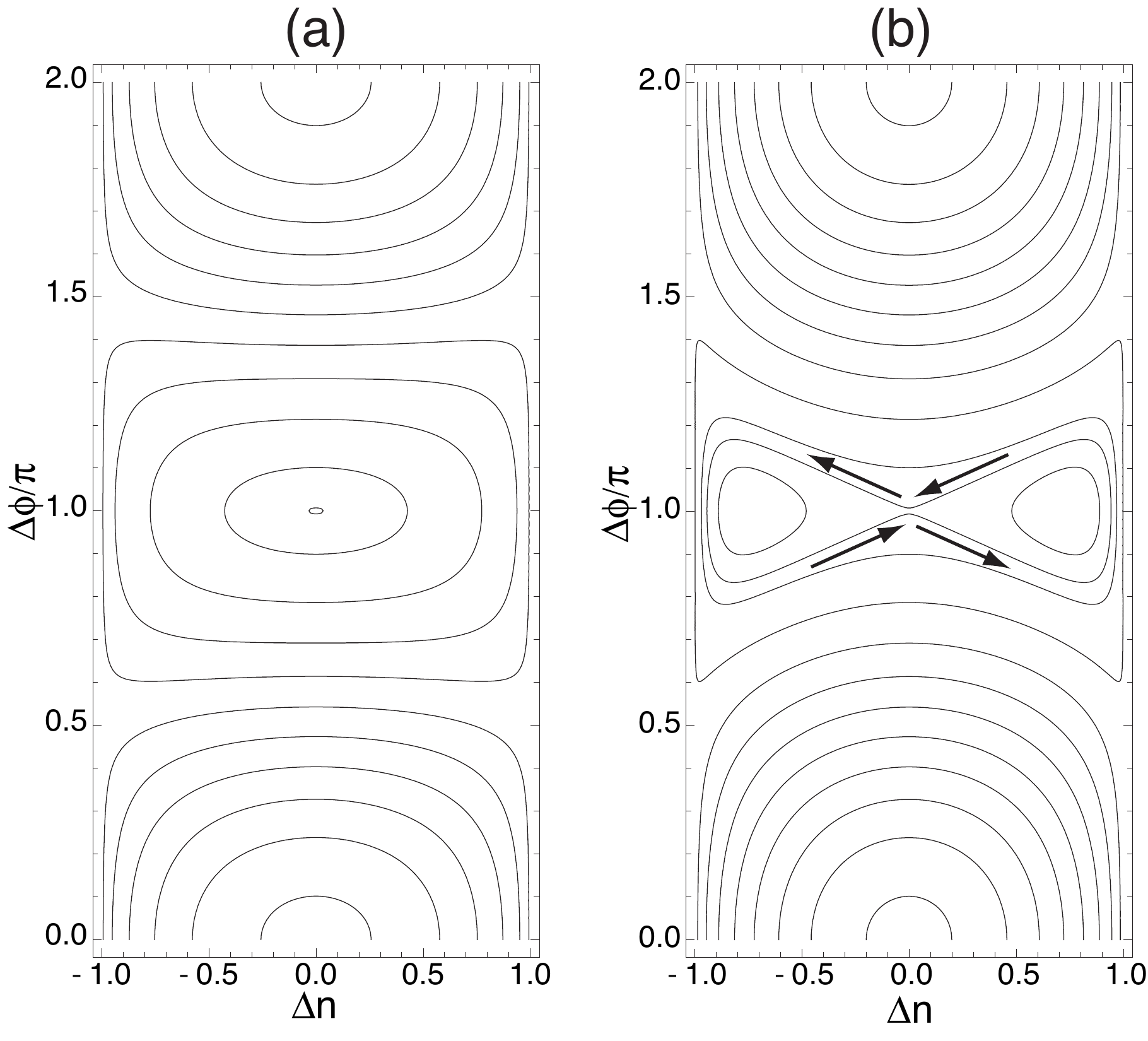}
\vspace{-20pt}
\end{center}
\caption{Contour plots for the Hamiltonian in the case of two sites only, in the ($\Delta n$,$\Delta\phi$) plane, for the interactions strengths (a) $\chi/\delta = 0.5$ and (b) 1.5. The arrows in panel (b) indicate the direction of flow.}
\vspace{-10pt}
\label{CONTOURPLOTS}    
\end{figure}

The Hamiltonian~\eq{TWOSITEH} is also a constant of the motion, which makes the two-site system integrable. The motion in phase space is along constant-energy curves in the ($\Delta n,\Delta\phi$) plane. Accordingly, in Fig.~\ref{CONTOURPLOTS} we draw constant-energy contours in this plane for $\chi/\delta=0.5$ (a), and $\chi/\delta=1.5$ (b). In these figures the $\Delta\phi$  axis runs from 0 to $2\pi$, so that the potentially unstable steady state is at the center of the contour plot; the unconditionally stable steady state with $P=0$, or $\Delta n=0$, $\Delta\phi=0$ is split between the middle of the lower and upper edges of the contour plot. For $\chi/\delta=0.5$ the potentially unstable steady state is an elliptic fixed point, and time evolution starting in its vicinity takes the system periodically around the stationary state. At  $\chi/\delta=1$ the elliptic fixed point bifurcates, and for $\chi/\delta=1.5$ there is a homoclinic point plus directions along which the system either recedes from the homoclinic point exponentially in time, or approaches the homoclinic point according to a negative-exponential law. The directions of the flow are indicated in panel~(b). Thus, starting in the vicinity of what used to be the potentially unstable steady state, the system takes off in an unstable direction and goes around one of the bifurcated elliptic fixed points. Depending on where the system starts out, it may have to loop around the second new elliptic fixed point, too, before returning to the original state. 

Multisite systems share the property of the two-site system that, starting from noise in the neighborhood of the unstable steady state, they may execute a large excursion out of the vicinity of the steady state but then return. We surmise that the mechanism of the recurrences in multisite and two-site systems is basically the same. As the interaction strength increases, the would-be unstable fixed point bifurcates and generates a homoclinic point along with two elliptic fixed points; actually, two overlapping homoclinic points for the modes $Q=\pm1$. Starting from noise, the system is propelled along a newly created unstable direction, which is the same for both modes $Q=\pm1$. A trajectory starting from the homoclinic point in an unstable direction returns to the homoclinic point along a stable direction. The multisite system appears to stay in the vicinity of such a homoclinic orbit.

Depending on the motion in the neighborhood of the homoclinic orbit due to the initial noise, the multisite system does not return to exactly where it started from. This may account for some of the variations in the period. Nonetheless, for small initial noise, the system should spend most of its time either receding from the homoclinic point, or approaching it. Accordingly, we have verified by numerical experiments that the time scale of the instability  $\tau= 1/|\Im\omega|$ is the dominant time scale for the recurrences.

Although the two-site model offers a neat overall explanation of the pulsating instability, theory of nonlinear systems should offer a more fine-graded description. We could have here a $q$-breather~\cite{FLA05q}, or an oscillatory instability~\cite{JOH99,FLA05} that takes the lattice between the unstable plane wave and a breather. The difference between the time for the first peak and the intervals between the subsequent peaks could be because the system initially radiates the energy that does not participate in the oscillations into the background plane wave modes. The details remain to be investigated.

\begin{figure}
\begin{center}
\includegraphics[width=8.0cm]{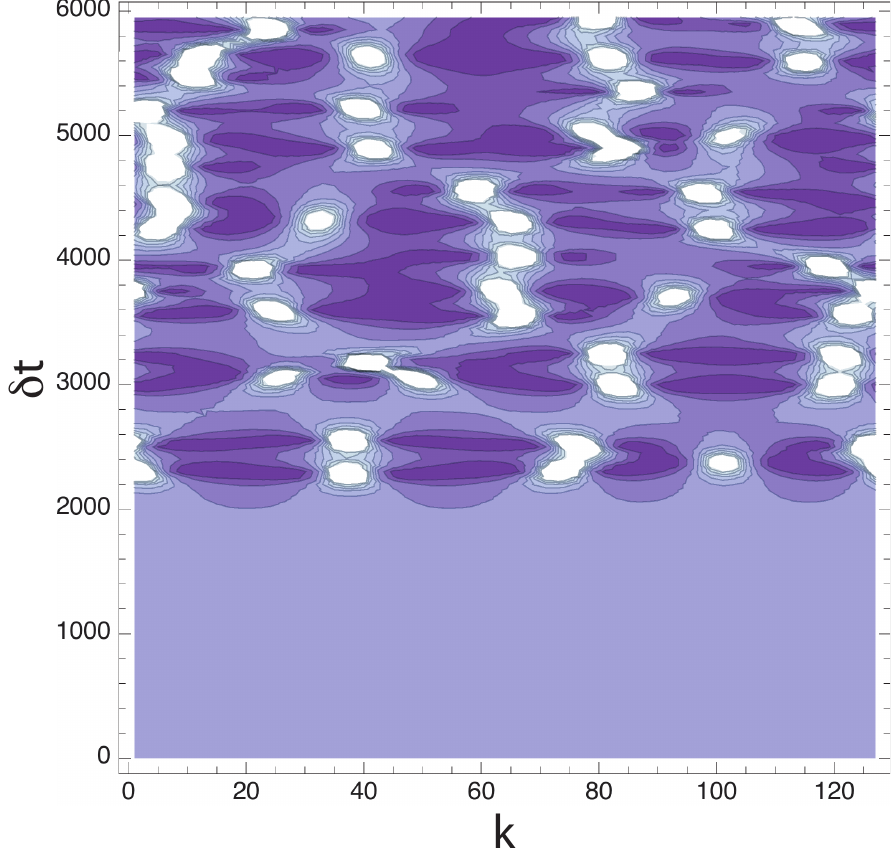}
\vspace{-20pt}
\end{center}
\caption{(color online). Density plot for the populations $|\alpha_k|^2$ as a function of time for the parameters $N=128$, $\chi/\delta = 0.015$, and $\xi=10^{-4}$; the darker the area, the smaller the value.}
\vspace{-15pt}
\label{GRAYSCALE}    
\end{figure}

For a lattice with a large number of sites it may be difficult to realize the one-peak condition~\eq{PEAKCONDITION} experimentally, but there is some leeway: the noise may seed several approximately independent pulsating peaks in the lattice. Figure~\ref{GRAYSCALE} shows an example. Here we have $N=128$ lattice sites, the interaction strength is $\chi/\delta = 0.015$, and initial random noise on the $p=\pi$ mode is specified by $\xi=10^{-4}$. The populations $|\alpha_k|^2$ are presented on a density plot as a function of time, the darkest color indicating  $|\alpha_k|^2\simeq0$ and white the largest populations. The right edge of the plot wraps around to the left edge by virtue of the periodic boundary conditions. The peaks represented by the white spots are not independent, they move around, join, and split, but most of the time there clearly are four pulsating peaks. For the parameters of Fig.~\ref{GRAYSCALE} the width of the peaks from Eq.~\eq{PEAKWIDTH} is $\ell\sim13$, so that four peaks and a peak's width of average spacing between them would require about 100 lattice sites, five peaks 130. The peak-width argument agrees with the numerical observations.

Flow states with $p\simeq\pi$ might be prepared, e.g., by accelerating the lattice~\cite{EIE04}, and there is also an alternative way. Namely, if $\alpha_k(t)$ are the solution to the equation of motion~\eq{EQM} for the parameters ($\delta,\chi$), then  $(-1)^k\alpha^*_k(t)$ are the solution for the parameters ($\delta,-\chi$). The ground state $p=0$ for a lattice with repulsive interactions behaves like the state $p=\pi$ after the sign of the interactions is suddenly flipped. This type of an experiment has been carried out using a Feshbach resonance~\cite{STR02}, albeit in a nonlattice gas and with strong interactions, giving rise to a chain of stable solitons. However, the pulsating instability occurs at weak interactions, and the corresponding time scales may present severe technical and fundamental challenges. For instance, quantum fluctuations~\cite{RUO05} may have to be considered. Given that the pulsations are a result of near-integrability, we speculate that similar phenomena take place in lattice with hard-wall boundary conditions and even in the presence of soft trapping in the direction of the lattice.

There is more to the instability of an unstable flow in an optical lattice than devolution into irregular behavior or soliton formation. In the case when the nonlinearity is weak,  the instability may lead to quasiperiodic pulsation where the atoms first gather and then return to the initial unstable state. With an increasing number of lattice sites, one may also see multiple peaks in atom populations oscillating in a similar way. A picture based on the properties of the phase space of a two-site lattice qualitatively explains the quasiperiodic revivals as a remnant of the integrability of the two-site system. Exceedingly simple arguments give a semi-quantitative description useful for the assessment of the feasibility, and for the design, of real experiments.

This work is supported in part by NSF (PHY-0750668).

\end{document}